\documentstyle[12pt]{article}

\begin{document}
\textwidth 17cm
\textheight 20 cm
\begin{flushright}
VAND-TH-94-9\\
May 1994\\
\end{flushright}
\begin{center}
\LARGE
{\bf Energy Density of Non-Minimally Coupled Scalar
Field Cosmologies}
\end{center}
\normalsize
\bigskip
\begin{center}

{\large David Hochberg\footnote{Present address:
 Departamento de F\'isica Te\'orica,
Universidad Aut\'onoma de Madrid, Cantoblanco, 28035 Madrid, Spain}  }
 and {\large Thomas W. Kephart}\\
{\sl Department of Physics and Astronomy, Vanderbilt University\\
Nashville, TN  37235, USA}
\end{center}
\vspace{2cm}
\centerline{ABSTRACT}
Scalar fields coupled to gravity via
$\xi R {\Phi}^2$ in arbitrary Friedmann-Robertson-Walker
backgrounds can be represented by an effective flat space
field theory. We derive an expression
for the scalar energy density where
the effective scalar mass becomes an explicit function
of $\xi$ and the scale factor. The scalar quartic self-coupling
gets shifted and can vanish for a particular choice of $\xi$.
Gravitationally induced symmetry breaking and de-stabilization
are possible in this theory.

\vfill\eject

\noindent
{\bf I. Introduction}

The study of classical and quantum fields in curved background
spacetimes is by now a well established first step
towards incorporating the influence of gravitation on both
microscopic and large-scale physics. A major interest in this
subject was spawned in the mid-seventies which saw a tremendous
effort devoted towards the development of rigorous techniques for
computing renormalized stress-energy tensors for quantum fields
in various background geometries [1]. The knowledge of such
tensors serves as the starting point for computing the back-reaction
on the metric, an application of which is the semi-classical theory
of black holes [2]. Apart from the stress-tensor renormalization
program, cosmology has provided a strong motivation for delving into
curved space field theory, most notably within the context of
inflationary models [3] and the large-scale structure problem [4].
In both these areas, scalar fields play a decisive role. The inflationary
epoch is believed to be driven by scalar field dynamics, usually via
symmetry
breaking. In the structure formation problem, (gauge-invariant)
density perturbations are resolved into tensor, vector and scalar
components. It is the dynamical equations for the scalar
perturbation that reveal
information on the growth and decay of density fluctuations [5].

In this paper, we take our cue from cosmology and consider the
problem of scalar fields in Friedmann-Robertson-Walker
 (FRW) spacetimes.  For our main result, we
demonstrate that a scalar in an
arbitrary expanding background, coupled with arbitrary strength
to the scalar curvature $(\xi R {\Phi}^2)$,
is equivalent to a scalar with a time-dependent mass in a
flat, Minkowski background. Moreover, the scalar quartic
self-coupling in the effective theory
is proportional to the FRW scalar quartic coupling, but rescaled
by a term involving $\xi$.
The equivalence between curved-space and flat-space scalar field
theory
is established by computing the scalar field
Hamiltonian, as this reveals the harmonic oscillator structure of
quantum fields in FRW spaces, and allows us to make our
curved space - to - flat space correspondences [6].
The benefit of the
effective flat space representation will become apparent when we
discuss implications for symmetry breaking in the early Universe
and constraints on the non-minimal coupling from stability
criteria.

Before launching into our analysis let us pause to briefly mention
previous related work.  Non-minimally  coupled scalar field
cosmologies have already been studied extensively by a number of
authors.  Two of the major topics that have been investigated
are: (i) the cosmic no hair conjecture relating to the smoothing
of anisotropies in the early Universe during inflation [7]
and (ii) how nonminimal coupling
affects inflation;
it has been shown how variations in  $\xi$
may or may not allow
chaotic or power law inflation in various cosmologies [8]. Here
we will address neither of these topics as our major interest is in the energy
spectrum of the scalar fields.
Our derivation of an energy spectrum for the scalar fields utilizes a
Bogolyubov transformation
and generates a conformally flat effective theory
in the process.  A flat space action
can also be derived directly via a conformal transformation, but
is less efficient for finding the energy spectrum.
The lowest order effective action for gauge
invariant cosmological perturbations is that of
a real scalar with a time dependent mass in a flat
background, ref. [23].
However, in addition to being a perturbative result, both
the self-coupling and nonminimal coupling to gravity are not considered in
those works. By contrast, our results are exact, and valid for
arbitrary $\xi$ and $\lambda$, but do not involve gauge invariant
perturbations about the FRW background.

\noindent
{\bf II. Energy and Effective Mass }

In treating the energy content of fields over curved backgrounds, a careful
and consistent choice must be made as to {\it which} of various and
possible field functionals actually represents the true, physical energy.
For example, the canonical Hamiltonian, defined as the spatial integral
of the Legendre-transformed Lagrange density, provides one possible
candidate for an energy-like expression. Another candidate is furnished
by the spatial integral of the pure timelike components of the stress-energy
tensor, obtained from functionally differentiating the Lagrange density
with respect to the background metric. In Minkowski space, there is no
difference between the energies defined by one or the other of these two
constructs. However, it is known that in expanding spacetimes, energy
functionals derived via the canonical Hamiltonian and stress-energy tensor
are generally {\it not} identical [9]. Moreover, in such backgrounds,
inequivalent Hamiltonian
energy expressions can be generated by means of sucessive canonical
transformations. To avoid this ambiguity, we will define the energy via
the stress-energy tensor. The latter has, after all, a direct physical
significance as the source of the gravitational field.

Specializing to the case of a real scalar field,
the variation of the action with respect to
the metric results in a stress tensor:
\begin{eqnarray*}
 T_{\mu \nu} (\Phi) & = & {2 \over {\sqrt{-g}}} {{\delta A} \over
 {\delta g^{\mu \nu}}}  =  (2\xi - {1 \over 2}) g_{\mu \nu} (g^{\alpha
\beta} \partial_{\alpha} \Phi \partial_{\beta} \Phi) - (2\xi -1)
\partial_{\mu} \Phi \partial_{\nu} \Phi
  -  2\xi \Phi \nabla_{\mu} \nabla_{\nu} \Phi \\
& + & 2\xi g_{\mu \nu} \Phi \Box \Phi -
\xi(R_{\mu \nu} - {1 \over 2}g_{\mu \nu} R){\Phi}^2
+ {1 \over 2}m^2 {\Phi}^2 g_{\mu \nu} + {{\lambda} \over 8}{\Phi}^4
g_{\mu \nu} ,
\end{eqnarray*}
$$\eqno(1)$$
where $\nabla_{\mu}$ is the covariant derivative and
$\Box = {1 \over {\sqrt{-g}}} \partial_{\mu}(\sqrt{-g} g^{\mu \nu}
\partial_{\nu})$.
We have taken
the most general action for a real scalar field in an arbitrary curved
spacetime,
$$A = \int d^4x\, \sqrt{-g}\, (g^{\mu \nu} \partial_{\mu} \Phi
 \partial_{\nu} \Phi - \xi {\Phi}^2 R -
V(\Phi)),\eqno(2)$$
where $V(\Phi) = m^2 {\Phi}^2 + {{\lambda} \over 4} {\Phi}^4$ [10].
This includes a possible nonminimal coupling to the metric,
where $\xi$ is a real
coupling constant and $R$ is the Ricci scalar curvature.
This coupling can assume any real value; when $\xi = 0$, the scalar is
said to be minimally coupled,
when $\xi = 1/6$, the scalar is conformally coupled.

As our primary interest is in cosmological applications,
we consider homogeneous
isotropic Friedmann universes with zero spatial curvature [11], having
as line element
$ds^2 = dt^2 - a^2(t)\, d{\bf x \cdot}d{\bf x},$
where $a(t)$ is the scale factor. It is convenient to express
the metric in terms of conformal coordinates $(\eta, {\bf x})$ where
$\eta(t) = \int dt'/a(t')$ is conformal time and $C(\eta) = a^2(t)$
is the conformal scale factor. Then, the metric components are
simply
$g_{\mu \nu} = C(\eta)\, {\rm diag}(1,-1,-1,-1).$
In these coordinates, the physical energy
density is ${\cal E} = T_0^{0}$,
where
\begin{eqnarray*}
 T_{00}  =  {1 \over 2}{\dot \Phi}^2 & + & ({1 \over 2} -
2\xi) {\bf \nabla}\Phi {\cdot} {\bf \nabla}\Phi
- 2\xi \Phi({\ddot \Phi}
- {{\dot C} \over {2C}} {\dot \Phi})\\
& + & \left[ ({1 \over 2} - 2\xi)m^2 C - \xi(R_{00} + (2\xi - {1 \over 2})
CR) \right] {\Phi}^2 + \lambda ({1 \over 8} - \xi) C {\Phi}^4,\, (3)
\end{eqnarray*}
 ${\bf \nabla}$ denotes the ordinary flat gradient, the overdot
represents ${\partial} \over {\partial \eta}$, and we have used
the equation of motion
$$(\Box + m^2 + \xi\,R + {{\lambda} \over 2}{\Phi}^2 )\Phi = 0,\eqno(4)$$
in arriving at (3). The classical energy is obtained by
integrating the energy density (in units where $\hbar = 1$)
$$E = \int d^3{\bf x}\, \sqrt{-{}^{(3)} g}\, T_0^0 (\Phi),\eqno(5)$$
over the spatial sections, while using the determinant of the three-metric
of those subspaces.

The passage to the quantum mechanical energy operator is
achieved in the standard way by expanding the classical solutions
of (4) in terms of a complete set of mode functions and imposing
canonical commutation relations on the generalized Fourier
coefficients [1]. The scalar field equation (for $\lambda = 0$)
is solved with the
ansatz
$$u_k(\eta, {\bf x}) = N_k \,C^{-1/2} \, f_k(\eta)\,
{{e^{i\bf k \cdot x}} \over {(2\pi)^{3/2}}}, \eqno(6)$$
where $f_k$ is a solution of
$${\ddot f_k} + \left[ k^2 + m^2 C(\eta) + (\xi - {1 \over 6})
R(\eta)C(\eta) \right] f_k = 0,\eqno(7)$$
and $k = |{\bf k}|$.
The associated mode expansion for $\Phi$ is given by
$$\Phi = \int {{d^3{\bf k}}  \over {(2\pi)^{3/2}}}\,
 (a({\bf k}) u_k + a^{\dagger}({\bf k}) u^{\ast}_k),\eqno(8)$$
where
$[a({\bf k}), a^{\dagger}({\bf k'})] = \delta^3({\bf k - k'})$, and
the other commutators vanish identically. The orthonormality
of the mode functions $u_k$ with respect to the (curved space)
inner product (for a discussion of normalization see [1];
the reduction to flat space is given below)
$$(F,G) = -i C(\eta) \int d^3{\bf x}\, (F {{\stackrel{\leftrightarrow}{
\partial}} \over {\partial \eta}} G^{\ast} ) \eqno(9)$$
is expressed by the conditions
$(u_k,u_p) = \delta^3({\bf k - p})$,
$(u^{\ast}_k, u^{\ast}_p) = -\delta^3({\bf k - p}),$
and $(u_k, u^{\ast}_p) = 0.$
Inserting the mode solutions (6) into (9)
fixes the $k^{th}$ mode normalization constant
$$N^2_k\,(f_k {\dot f^{\ast}_k} - {\dot f_k} f^{\ast}_k) = i.
\eqno(10)$$
The polar representation of the function $f_k$ is the starting point
for a WKB analysis of the approximate solutions of (7) [12] and
is also useful for making {\it exact} comparisons
between curved-space results in
field theory and their Minkowski space counterparts. In this representation
we write
$$f_k = A_k(\eta)\, e^{-i S_k(\eta)},\eqno(11)$$
where $A_k, S_k$ are real amplitude and phase functions. Substituting
this into (10) yields the important relation
$$N^2_k A^2_k\, {\dot S_k} = {1 \over 2},\eqno(12)$$
which we will have occasion to use later on.
For example, in the Minkowski limit, the solutions of (7) are
$f_k \sim e^{\pm \imath \omega_k \eta}$ with $A_k = 1$ (without
loss of generality)
and constant phase,
${\dot S_k} = \omega_k = (k^2 + m^2)^{1/2}$. In this limit, the
mode normalization (12) reduces to
$N_k = {1 \over {\sqrt{2\omega_k}}}$, and $E_k = \omega_k$,
which one immediately recognizes
as the standard results for flat space field theory [13].

Inserting the field operator (8) into (5) and (3) gives the
energy operator. Unlike the familiar flat space
result of the last paragraph, this
involves second derivatives of the mode functions in (6),
leading to
intermediate steps which are straightforward
but rather lengthy.
We turn first to the quadratic part, for which we obtain
\begin{eqnarray*}
E/{C^{1/2}} = & \int& {{d^3{\bf k}} \over {(2\pi)^{3/2}}}
N^2_k A^2_k \, ( Z_k\,  a({\bf k}) a^{\dagger}({\bf k})  +
 Z^{\ast}_k \,a^{\dagger}({\bf k}) a({\bf k})  \\
 & + &   W_k \,a({\bf k}) a({\bf -k}) +
W^{\ast}_k \, a^{\dagger}({\bf k})
  a^{\dagger}({\bf -k}) ) \,\,\,  (13)
\end{eqnarray*}
where
$$Z_k = M^2 + {1 \over 2}(\omega^2_k + Y^2_k) + ({1 \over 2} - 2\xi)k^2
- 2\xi \left[ {\cal F^{\ast}}_k - {{\dot C} \over {2C}}
(i\omega_k + Y_k) \right],
\eqno(14)$$
and
$$W_k \,e^{2iS_k} = M^2 + {1 \over 2}(-i\omega_k + Y_k)^2 + ({1 \over 2} -
2\xi)
k^2 - 2\xi \left[ {\cal F}_k -{{\dot C} \over {2C}}
(-i\omega_k + Y_k) \right],
\eqno(15)$$
are complex functions of $\eta$. Here, $M^2$ is a term combining the mass
with a curvature dependent function [14],
$$M^2 = ({1 \over 2} - 2\xi) m^2 C - \xi (R_{00} + (2\xi - {1 \over 2})
CR).\eqno(16)$$
In computing the derivatives of the $u_k$, we have found it useful
to define the following quantities
$Y_k = {{\dot A_k} \over {A_k}} - {{\dot C} \over {2C}},$
and
${\cal F}_k = (-i{\ddot S_k} + {\dot Y_k}) + ( -i{\dot S_k} + Y_k)^2$
where
$\omega_k \equiv {\dot S_k}$
is the instantaneous frequency of the $k^{th}$ mode [15].

The energy operator in (13) can be brought to diagonal form by means
of a Bogolyubov transformation  (a canonical transformation)
[16]
$$a({\bf k}) = \alpha_k\, b({\bf k}) + \beta_k\, b^{\dagger}({\bf -k}),\eqno
(17a)$$
$$a^{\dagger}({\bf k}) = \alpha^{\ast}_k\, b^{\dagger}({\bf k})
 + \beta^{\ast}_k \, b({\bf -k}),\eqno(17b)$$
where the new operators $b({\bf k})$ obey canonical commutation
relations, viz. $[b({\bf k}), b^{\dagger}({\bf k'})] = \delta^3(
{\bf k - k'})$, etc., provided the Bogolyubov coefficients
satisfy $|\alpha_k|^2 - |\beta_k|^2 = 1$. For the case at hand, we can
parametrize this transformation by a single real angle by taking
$$\alpha_k = \cosh(\theta_k) = {1 \over {\sqrt{1 - L_k^2}}}, \qquad
{\rm and} \qquad
\beta_k = \sinh(\theta_k) = {{L_k} \over {\sqrt{1 - L^2_k}}},
\eqno(18)$$
where $L_k = L_{-k}$ is a real function [17]. Applying the
transformation (17) to (13), and absorbing the phase factor
$e^{2iS_k}$ into the definition of $W_k$,
we find that
$$ E \rightarrow E_{number} + E_{squeeze},\eqno(19)$$
where ($\Re$ denotes the real part)
$$E_{number} = \int d^3{\bf k}\, {{2N^2_k A^2_k C^{-1/2}}
\over {(1 - L^2_k)}}\,
\left( (1 + L^2_k)\, \Re Z_k  + 2 L_k\, \Re W_k
\right) b^{\dagger}({\bf k}) b({\bf k}) ,\eqno(20)$$
and
$$E_{squeeze} = \int d^3{\bf k}\, {{N^2_k A^2_k C^{-1/2}} \over {(1 -
L^2_k)}}\,
\left( 2 L_k\, \Re Z_k + W_k + L^2_k W^{\ast}_k \right)
 b({\bf k})b({\bf -k}) + hc.\eqno(21)$$
The `squeeze' term is so called because it is identical in form to
multi-mode squeeze operators familiar from quantum optics [18] and
leads to the phenomenon of gravitational squeezing as discussed
in [19,20]. The
other term has the familiar form of a weighted sum of number
operators (after normal-ordering with respect to the $b$-vacuum [21]).
Thus, diagonalization is achieved for the unique choice of time-dependent
transformation angle corresponding to
$$L_k = {{-\Re Z_k + ((\Re Z_k)^2 - |W_k|^2)^{1/2}} \over
{W^{\ast}_k}}.\eqno(22)$$
For this choice, $E_{squeeze} = 0$ and $E = E_{diag} \equiv E_{number}$.
This solution is fixed by the requirement that $L_k \rightarrow 0$
as the background field is shut off, i.e., as $a(t) \rightarrow  1$.
In this limit, the angle $\theta_k \rightarrow 0$ , $\alpha _k
\rightarrow 1$ and $\beta_k \rightarrow 0$, as expected.

While the coefficient in (20) with $L_k$ given in (22) gives
the correct energy eigenvalues, the resultant expression
admits a tremendous algebraic simplification, though this fact
is by no means a-priori obvious. To see how this comes about, we
write the integrand of (13) as a 4-by-4 matrix for each mode $k$
$${\cal E}_k = N^2_kA^2_k C^{-1/2} \left( \begin{array}{cccc}
                      0 & Z_k & W_k & 0 \\
                 Z^{\ast}_k & 0 & 0 & W^{\ast}_k\\
                     W_k & 0 & 0 & Z_k \\
                      0 & W^{\ast}_k & Z^{\ast}_k & 0 \\
                     \end{array}
                    \, \right),\eqno(23)$$
where the rows and columns of the array are labeled in the
sequence $(a_k, a^{\dagger}_k, a_{-k}, a^{\dagger}_{-k})$.
The determinant of this matrix is easily computed:
$$\det {\cal E}_k = (N^2_kA^2_kC^{-1/2})^4\, ( |Z_k|^2 - |W_k|^2)^2.\eqno(24)$$
On the other hand, as we have just demonstrated, the Bogolyubov
transformation brings the energy operator into the diagonal (with
respect to the number-operator basis) [22]
form given by
$${\cal E}_k = \left( \begin{array}{cccc}
               0 & E_k & 0 & 0\\
              E_k & 0 & 0 & 0 \\
               0 & 0 & 0 & E_k \\
               0 & 0 & E_k & 0 \\
              \end{array}
               \,\right).\eqno(25)$$

The canonical transformation (17a,b)
is just a symplectic transformation which
preserves the quadratric form $|\alpha|^2 - |\beta|^2 = 1$.
As the spectrum is invariant under this change of basis, we
must have (using $E_k$ to denote both the operator and its
eigenvalue, without loss of generality)
$$E_k =  N^2_k A^2_k C^{-1/2} \,| |Z_k|^2 - |W_k|^2 |^{1/2}.\eqno(26)$$
The calculation of the energy levels is now straightforward and
follows from substituting the expressions for $Z_k$ and $W_k$
into (26). The elimination of the second derivatives of the
amplitude and phase functions which appear there is effected
by means of the two identities
$$A_k\,{\ddot S_k} + 2{\dot A_k}\,{\dot S_k} = 0,\eqno(27a)$$
and
$$\left( {{\ddot A_k} \over {A_k}} \right)
- ({\dot S_k})^2 + (k^2 + C[m^2 + (\xi - 1/6)R]) = 0,\eqno(27b)$$
which result from substituting the polar functions (11) into the
differential equation (7) [12]. Note the first equation can be
integrated immediately to give
$$ A_k \,\sqrt{\dot S_k} = b,\eqno(28)$$
for some constant $b$, and gives independent confirmation of the
Wronskian constraint in (10). Comparison with (12) implies
$b = {1 \over {\sqrt{2} N_k}}$. After some additional algebra
we find that
$$|Z_k|^2 - |W_k|^2 = \omega^2_k \, (m^2 C + k^2 + {3 \over 2}
\xi ({{\dot C} \over {C}})^2 ),\eqno(29)$$
which together with (12) and (26) yields the quadratic
energy spectrum
$$E_k =  \left( m^2  + k_{phys}^2 + 6 \xi (H(t)/c)^2 \right)^{1/2}
,\eqno(30)$$
which is our key result. Note that (30) is an exact result for a
``free" scalar $(\lambda = 0)$, but nonminimally coupled to
gravity $(\xi \neq 0)$ [23].
The physical wave vector
${\bf k}_{phys} = {\bf k}/a(t)$ reflects the Doppler spread in
the wavelength caused by the expansion/contraction of the background
spacetime, and $H = {1 \over a}{{da} \over {dt}}$ is the Hubble constant.
In the diagonal basis
introduced in (17), the energy is given by
$$E = \sum_{\bf k} E_k \, b^{\dagger}({\bf k}) b({\bf k}) + E_{int}(
\lambda),\eqno(31)$$
where $E_{int}$ depends on the self-coupling. This result demonstrates that
a free scalar field in a FRW
background is equivalent to a system of
harmonic oscillators in flat space, with the proviso
that the oscillators have an
effective, time-dependent, mass given by
$$m^2_{eff}(\xi,t) \equiv m^2 + 6 \xi (H/c)^2.\eqno(32)$$

We can write down an action leading to the same energy operator (31)
if we identify the mass parameter appearing there with (32).
The interacting part can be incorporated immediately since the same
transformation which diagonalized the quadratic sector keeps the
quartic terms invariant (and these do not mix with the quadratic
part). The effective quartic coupling can then be read off
immediately from (3). The resultant effective field theory
is given by [24]
$$A_{eff} = \int d^4x\, (({\partial_t \phi})^2 - ({{\bf \nabla}_{phys}\phi})^2
 - m^2_{eff} {\phi}^2 - {{\lambda_{eff}} \over 4}
{\phi}^4),\eqno(33)$$
where $\lambda_{eff} = (1 - 8\xi) \lambda$, and $\phi$
admits an expansion similar to (8) (when $\lambda_{eff} = 0$)
with $a({\bf k})$ replaced by $b({\bf k})$, and with different
mode functions.

In passing, we comment briefly on the properties of a theory with an
effective action given by (33), which has the $Z_2$ symmetry
$(\phi \rightarrow -\phi)$ [25]. For $\lambda_{eff} > 0$ the
flat space theory undergoes
spontaneous symmetry breaking (restoration)
for $m^2 < 0$ ($m^2 > 0$) and $\xi > 0$ ($\xi < 0$) at a time
${\hat t}$ when [26]
$$|m^2| = 6|\xi| ({ {H({\hat t})} \over c})^2,$$
that is, when the Compton wavelength of the scalar is roughly
the horizon size.

For $\lambda_{eff} < 0$ the flat space scalar
theory is unstable but asymptotically
free [27]. Finally, note there are two ways that (33) gives
the action for a free flat space theory: when $\lambda = 0$ or
when $\xi = 1/8$. Futhermore, for $\lambda > 0$, a conformally
coupled scalar theory ($\xi = 1/6$) has $\lambda_{eff} =
-{\lambda}/3$, and hence the theory is unstable.

\noindent
{\bf III. Conclusions}

The purpose of obtaining the results presented here has been to improve our
understanding of the longest wavelength modes in cosmology, specifically, long
wavelength (horizon size or longer) scalar density perturbations.
We have demonstrated that considerable care must be taken in the
interpretation of  such modes (beyond the standard considerations of gauge
dependence).  Previous work [12,15] has shown that the relation $\lambda =
1/{\nu}$ holds only in flat space and in some special cosmologies.  Here, by
studying the energy spectrum of classical  scalar fields coupled to curvature,
we have learned that Einstein's relation $E = \omega$ (or $E =
\sqrt{{\omega}^2 + m^2}$ for massive scalar fields) no longer holds
in general, and what will be more important for the interpretation of the
physics of large scale structures: the energy is no longer related
in the standard way to the wavelength of the scalar fields that
are not minimally coupled in any cosmology where the Hubble parameter
is nonvanishing.  A challange for the future is to relate the results derived
here for the classical fundamental scalar fields $\Phi$ in  a homogeneous
cosmology to  the scalar density perturbation in a cosmological
background. It would also be of interest to explore vector and tensor
density perturbations in this approach [28].

This work was supported in part by DOE Grant No. DE-FG05-85ER40226.

\vspace{2cm}
\noindent
[1]. N.D. Birrell and P.C.W. Davies, {\it Quantum Fields in Curved
Space}, (Cambridge University Press, Cambridge, 1982).

\noindent
[2]. J.W. York, Jr., Phys. Rev. D{\bf 31}, 775 (1985); D. Hochberg and
T.W. Kephart, Phys. Rev. D{\bf 47}, 1465 (1993);
D. Hochberg, T.W. Kephart and J.W. York, Jr., Phys. Rev. D{\bf 48}, 497
(1993).

\noindent
[3]. For a review of inflationary cosmology, see, e.g.,
E.W. Kolb and M.S. Turner, {\it The Early Universe},
(Addison-Wesley, New York, 1990) Chap 8;\\
K. Olive , Phys. Rep. {\bf 190}, 307 (1990); A.D. Linde,
{\it Particle Physics and Inflationary Cosmology},
(Harwood, 1990), and references contained in these works.

\noindent
[4]. Extensive treatments of cosmological perturbation theory
may be found in H. Kodama and M. Sasaki, Prog. Theo. Phys. Suppl.
{\bf 78}, 1 (1984), and V.F. Mukhanov, R.H. Brandenberger and
H. A. Feldman, Phys. Rep. {\bf 215}, 203 (1992).

\noindent
[5]. S. Dodelson, L. Knox, and E. W.
Kolb, Phys. Rev. Lett.{\bf 72}, 3444 (1994).

\noindent
[6]. We wish to emphasize that our correspondence is a tree-level
result and the shift in the mass and coupling constant arises
from the interaction of $\Phi$ with the classical background
metric. This shift is generated from the transformation
between FRW and Minkowski spacetimes we carry out below. Additional
corrections to the mass and couplings can arise from quantum loop
effects, and these can be calculated along the lines presented
in L. Parker and D.J. Toms, Phys. Rev. D{\bf 31}, 2424 (1985).
These latter contributions are due to the fluctuations in $\Phi$
about a fixed curved background.

\noindent
[7]. The no-hair theorem has been extensively studied by,
R. M. Wald,  Phys. Rev. D{\bf 28}, 2118 (1983);
I. Moss and V. Sahni,  Phys. Lett. B{\bf 178}, 159 (1986);
L. G. Jensen and J. A. Stein-Schabes, Phys. Rev. D{\bf 35}, 1146 (1987);
K. Maeda, Phys. Rev. D{\bf 39}, 3159 (1989);
K. Maeda, J. A. Stein-Schabes, and T.Futamase,
Phys. Rev. D{\bf 39}, 2848 (1989);
J. Yokoyama and K. Maeda, Phys. Rev. D{\bf 41}, 1047 (1990);
A. L. Birkin and K. Maeda, Phys. Rev. D{\bf 44}, 1691 (1991);
Y. Kitada and K. Maeda, Phys. Rev. D{\bf 45}, 1416 (1992);
Y. Kitada and K. Maeda, Class. and Qm. Grav. {\bf 10}, 703 (1993);
H. Shinai and  K. Maeda, Phys. Rev. D{\bf 48}, 3910 (1993).

\noindent
[8]. Investigations of
inflation with non-minimally coupled scalar fields can be found in,
J. Barrow and M. S. Turner, Nature, {\bf 292}, 35 (1981);
G. Steigman and  M. S. Turner,  Phys. Lett. B{\bf 128}, 295 (1983);
T. Rothman and M. S. Madsen, Phys. Lett. B{\bf 159}, 256 (1985);
T. Futamase and K. Maeda, Phys. Rev. D{\bf 39}, 399 (1989);
T. Futamase, T. Rothman and R. Matzner, Phys. Rev. D{\bf 39}, 405 (1989).

\noindent
[9]. S.A. Fulling, Gen. Rel. Gravitation, {\bf 10}, 807 (1979).

\noindent
[10]. We have omitted cubic terms by imposing a $Z_2$
$(\Phi \rightarrow -\Phi)$
symmetry
for simplicity. However, this
does not affect our results nor conclusions. Our calculations
can easily accomodate complex scalars, for which of course, cubic
terms are strictly absent, by charge conjugation.

\noindent
[11]. The spatial curvature is negligible during the early stages
of expansion; it is to this epoch we address our calculations.

\noindent
[12]. D. Hochberg and T.W. Kephart, Phys. Rev. D{\bf 45}, 2706 (1992).

\noindent
[13]. B.J. Bjorken and S.D. Drell, {\it Relativistic Quantum Fields},
(McGraw-Hill, New York, 1965).

\noindent
[14]. Although $M^2$ appears as the coefficient of $\Phi^2$ in
$T_{00}$, it is not to be interpreted as the physical mass of the
scalar. Indeed, the background gravitational field mixes up the
positive and negative frequency components of $\Phi$ (this is the
origin of gravitational squeezing) and necessitates the diagonalization
of the resultant hamiltonian, which we carry out below. The situation
is analogous to that encountered in the standard model, where the
vacuum expectation value of the Higgs behaves like a classical
background field mixing up the components of the gauge fields.
This leads to a
nondiagonal mass matrix which must be diagonalized in order to
correctly identify the mass eigenstates of the model.

\noindent
[15]. D. Hochberg and T.W. Kephart, Phys. Rev. Lett. {\bf 66}, 2553 (1991).

\noindent
[16]. N. N. Bogolyubov, J. Phys. U.S.S.R., {\bf 11}, 23 (1947).

\noindent
[17]. The Bogolyubov coefficients are in general complex, but here
it can be shown that a real transformation achieves the desired
operator diagonalization.

\noindent
[18]. See, e.g., P. Meystre and M. Sargent, {\it Elements of Quantum
Optics}, (Springer, Berlin, 1990) ch. 16.

\noindent
[19]. L.P. Grishchuk and Y.V. Sidorov, Phys. Rev. D{\bf 42}, 3413 (1990).

\noindent
[20]. D. Hochberg and T. W. Kephart, Phys. Lett. B{\bf 268}, 377 (1991).

\noindent
[21]. For simplicity, we have dropped a c-number
term corresponding to the
renormalized zero-point vacuum fluctuations. Vacuum energy contributes
to the effective cosmological constant. Extensive treatment of stress-tensor
renormalization is given in Reference [1].

\noindent
[22]. Though (25) is not diagonal in the usual sense, it is diagonal
in the number-operator basis. A trivial re-ordering of the rows and
columns would put this array into standard diagonal form, but this
is not necessary.

\noindent
[23]. M. Sasaki, Prog. Theo. Phys.,
{\bf 76}, 1036 (1986) and V.F. Mukhanov, Zh. Eksp. Teor. Fiz. {\bf 94},
1 (1988).

\noindent
[24]. The gradient contains the scale factor: $\nabla_{phys} =
{1 \over a} \nabla$.

\noindent
[25]. Or, we could extend these results to say, a $U(1)$ theory, by
letting $\phi^2 \rightarrow (\phi \phi^{\ast})$, as
discussed in [9]; non-abelian symmetries are no more difficult to
handle.

\noindent
[26]. In general, 1-loop corrections give $m^2$ a temperature
(i.e., time) dependence $m^2 (T) = m^2(0)
+ O(1) \times \lambda_{eff}^2 T^2$; see
L. Dolan and R. Jackiw, Phys. Rev. D{\bf 9}, 3320 (1974);
S. Weinberg, Phys. Rev. D {\bf 9}, 3357 (1974). Here we
will ignore these corrections but in general,
these effects
must be taken into account, as can be easily done.

\noindent
[27]. G. `t Hooft, talk at the Marseilles meeting, 1972 (unpublished).

\noindent
[28]. One may expect (nongauged) classical fundamental vectors $v^{\mu}$
and tensors $t^{\mu}_{\nu}$ to couple to gravity via dimension four terms
of the form ${\xi}_1 R v^{\mu} v_{\mu} + {\xi}_2 R^{\mu}_{\nu}  v^{\nu}
v_{\mu} + {\xi}_3 R t^{\mu}_{\mu} + {\xi}_4  R^{\mu}_{\nu} t^{\nu}_{\mu}$
which in turn would contribute to vector and tensor density perturbations.

\end{document}